\documentclass[twocolumn,twocolappendix]{aastex631}
\usepackage{CJK}
\usepackage[toc,page]{appendix}
\received{2022-08-14}
\revised{2022-10-18}
\accepted{2022-10-27}
\published{2022-11-16, ApJL}

\newcommand\sdo{\emph{SDO}}
\newcommand\iris{\emph{IRIS}}
\newcommand\goes{\emph{GOES-15}}
\newcommand\kmps{\mbox{km s$^{-1}$}}
\newcommand\pcc{\mbox{cm$^{-3}$}}
\newcommand\pfc{\mbox{cm$^{-5}$}}

\newcommand\siiv{Si\,{\sc{iv}}} 
\newcommand\siivwav{1393.755 \AA}
\newcommand\siivall{\siiv{ }\siivwav}
\newcommand\mgii{Mg\,{\sc{ii}}} 

\def\papertitle{Spectroscopic and Imaging Observations
of Spatially Extended Magnetic Reconnection \\
in the Splitting of a Solar Filament Structure}
\def\paperkeywords{Solar magnetic reconnection (1504);
Solar filaments (1495); Solar filament eruptions (1981);
Ultraviolet spectroscopy (2284)}
\def\papersubject{Solar: magnetic reconnection; filament splitting; filament partial eruption}
\hypersetup{pdfauthor=Huidong Hu et al.,
pdftitle=\papertitle, pdfkeywords=\paperkeywords, pdfsubject=\papersubject}

\def\affnssc{State Key Laboratory of Space Weather, National Space Science Center,
    Chinese Academy of Sciences, Beijing 100190, China;
    \href{mailto:huhd@nssc.ac.cn}{huhd@nssc.ac.cn},
    \href{mailto:liuxying@swl.ac.cn}{liuxying@swl.ac.cn}}
\def\affucas{University of Chinese Academy of Sciences, Beijing 100049, China}
\def\affmps{Max Planck Institute for Solar System Research, G\"{o}ttingen D-37077, Germany}
\def\affnju{School of Astronomy and Space Science, Nanjing University, Nanjing 210023, China}
\def\affnjl{Key Laboratory for Modern Astronomy and Astrophysics (Nanjing University), Ministry of Education, Nanjing 210023, China}

\begin{document}
\begin{CJK*}{UTF8}{gbsn}
\title{\papertitle}
\author[0000-0001-8188-9013]{Huidong Hu (胡会东)}
\affiliation{\affnssc}
\author[0000-0002-3483-5909]{Ying D. Liu (刘颍)}
\affiliation{\affnssc}
\affiliation{\affucas}
\author[0000-0002-9270-6785]{Lakshmi Pradeep Chitta}
\affiliation{\affmps}
\author[0000-0001-9921-0937]{Hardi Peter}
\affiliation{\affmps}
\author[0000-0002-4978-4972]{Mingde Ding (丁明德)}
\affiliation{\affnju}
\affiliation{\affnjl}

\begin{abstract}
On the Sun,
Doppler shifts of bidirectional outflows
from the magnetic reconnection site
have been found only in confined regions
through spectroscopic observations.
Without spatially resolved spectroscopic observations
across an extended region,
the distribution of reconnection and its outflows
in the solar atmosphere cannot be made clear.
Magnetic reconnection is thought to cause the splitting of filament structures,
but unambiguous evidence has been elusive.
Here we report spectroscopic and {imaging analysis}
of a magnetic reconnection event on the Sun,
{using high resolution data
from the \emph{Interface Region Imaging Spectrograph (IRIS)}
and the \emph{Solar Dynamics Observatory (SDO)}}.
Our findings reveal that
the reconnection region extends to an unprecedented length
of no less than 14 000 km.
The reconnection splits a filament structure into two branches,
and the upper branch erupts eventually.
Doppler shifts indicate clear bidirectional outflows of $\sim$100 \kmps{},
which decelerate beyond the reconnection site.
{Differential-emission-measure analysis reveals that
in the reconnection region
the temperature reaches over 10 MK
and the thermal energy is much larger than the kinetic energy.}
{This Letter provides definite spectroscopic evidence
for the splitting of a solar filament by magnetic reconnection
in an extended region.}
\end{abstract}

\section{Introduction}\label{intro}
Magnetic reconnection is a process
that changes magnetic topology
and converts magnetic energy to plasma kinetic energy,
which exists in laboratory and astrophysical plasmas
\citep{ZweibelY2009ARAA}.
On the Sun, reconnection can contribute to
eruptions \citep[e.g.,][]{LinRB2004ApJ},
acceleration of energetic particles \citep[e.g.,][]{GoldsteinMA1986GeoRL},
and coronal heating \citep[e.g.,][]{AntolinPT2021NatAs}.

{One indicator of magnetic reconnection
is the Doppler effect of the reconnection outflows
\citep[e.g.,][]{InnesIA1997Natur,HongDL2016ApJ,PolitoGR2018ApJ}.
Blue- and redshifts of bidirectional reconnection outflows
are observed only in confined regions
\citep[e.g.,][]{ChiforYI2008AA,TianZP2018ApJ,OrtizHN2020AA}.
Spectroscopic observations
suggest that small-scale reconnection can}
drive small jets of multiple temperatures
\citep[e.g.,][]{ChiforYI2008AA,LiLN2018MNRAS}.
{Imaging and spectroscopy reveal
bidirectional outflows of reconnection that
can heat small pockets of cool plasma in the photosphere
\citep{PeterTC2014Sci}.
Doppler shifts of downward flows from the reconnection current sheet
are disclosed in the late stage of an eruptive flare on the solar limb
\citep{FrenchMv2020ApJ},
where the eruption is associated with a global coronal wave
\citep[e.g.,][]{HuHuidong2019ApJ}.}
Spatially resolved spectroscopic observations covering extended regions
in the solar atmosphere are rare,
and thus the distribution of reconnection
{outflows and thermal properties} on the Sun is unclear.

Models suggest that magnetic reconnection
can occur internally in a filament structure and
is associated with the splitting and/or partial eruption of the filament structure
\citep[e.g.,][]{GilbertHB2001ApJ,GibsonF2006ApJ,KliemTT2014ApJ}.
Imaging observations also indicate internal reconnection
related to the splitting and partial eruption of a filament
\citep[e.g.,][]{LiuGA2008ApJ,TripathiGQ2009AA,ChengKD2018ApJ}.
{In a ``double-decker'' system
that consists of two vertically separated filament branches,
magnetic reconnection between the two branches
can destabilize the upper branch and cause a partial eruption
\citep{LiuKT2012ApJ,KliemTT2014ApJ}.}
Doppler shifts of bidirectional outflows,
as a clear indicator of reconnection in solar filament splitting,
have not been detected in spectroscopic observations.
Therefore, definite evidence for reconnection
in the splitting of a solar filament has been elusive.

{In this Letter, we report a magnetic reconnection event
that causes the splitting of a solar filament structure,
based on spatially resolved spectroscopic data from
the \emph{Interface Region Imaging Spectrograph}
\citep[\iris,][]{PontieuTL2014SoPh}
and images from the \emph{Solar Dynamics Observatory}
\citep[\sdo,][]{PesnellTC2012SoPh}.}
An overview of the filament splitting
is described in Section \ref{observation}.
The spectroscopic results of the reconnection
are delivered in Section \ref{results}.
{The temperature and the energy
are estimated in Section \ref{dem}.}
The investigation is concluded and discussed in Section \ref{conclusions}.
The reconnection is unique in
that Doppler shifts of its bidirectional outflows
are observed in an unprecedented extended region on the Sun.
Our observations present clear spectroscopic evidence
for the splitting of a solar filament induced by magnetic reconnection.

\section{Observations and Overview}\label{observation}
\begin{figure*}[ht!]
\centering
\includegraphics[width=0.9\textwidth]{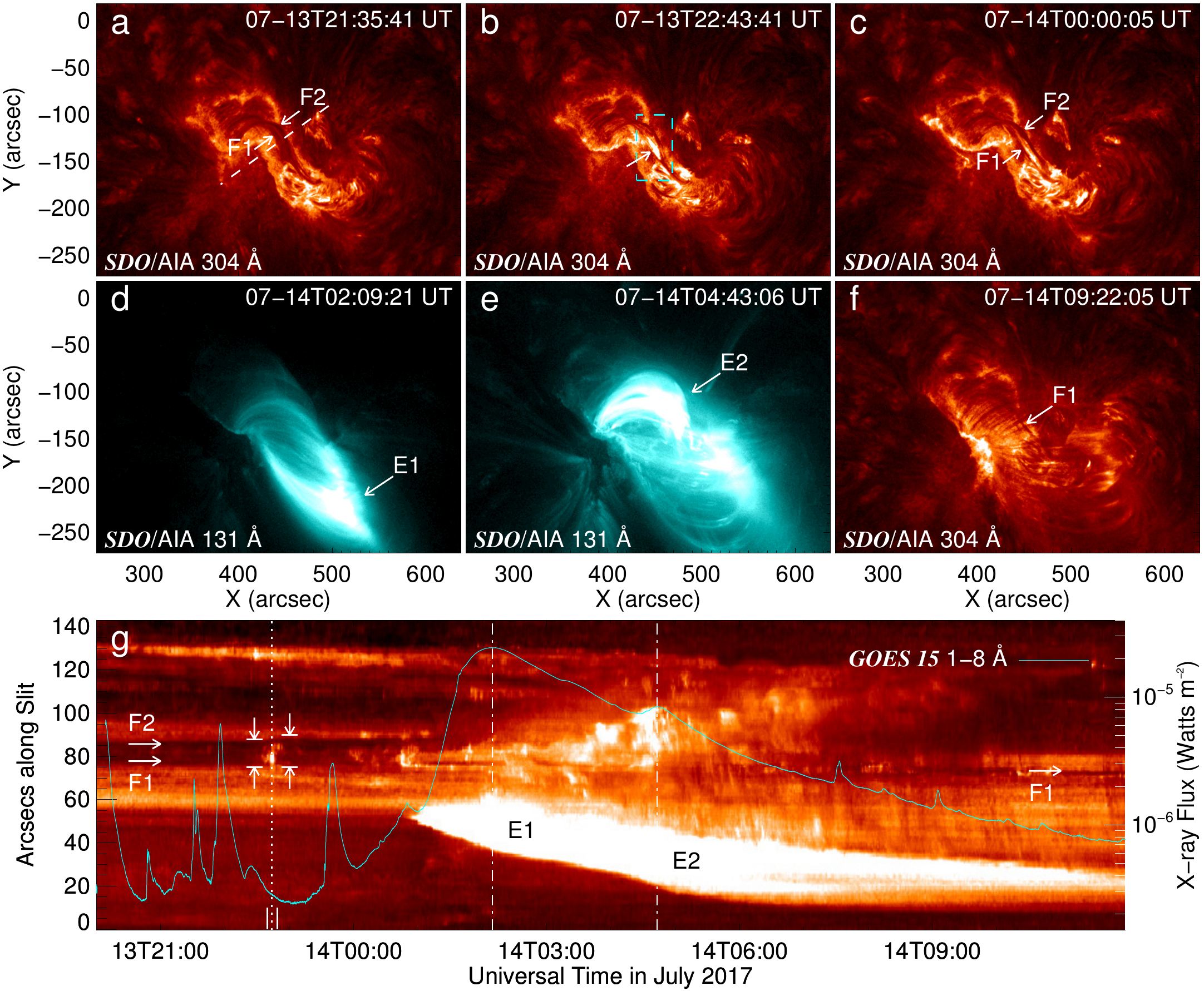}
\caption{\label{fig1}{Splitting and partial eruption of the filament structure.}
(a)-(f) Images of \sdo/AIA 304 \AA{} ((a)-(c), (f)) and 131 \AA{} ((d)-(e)).
(g) Time-distance profile created by stacking AIA 304 \AA{} images
along the dashed line in (a),
overlaid with \goes{} 1-8 \AA{} X-ray flux in cyan color.
{The field of views (FOVs) of (a)-(f) are the same,
whose coordinates are differentially rotated and are aligned to those of (b).}
``F1'' and ``F2''  in (a), (c), (f), and (g)
indicate the lower and upper filament branches, respectively.
In (b) the arrow marks a brightening region between the two branches signifying the splitting.
``E1'' and ``E2'' in (d)-(e) denote two flares associated with the partial eruption.
In (g): the dotted line on the left indicates the approximate time of the splitting,
and the nearby two pairs of arrows illustrate
the total width of the two filament branches before and after the splitting;
the two dash-dotted lines mark the times of the X-ray flux peaks of flares ``E1'' and ``E2''.
The dashed rectangle in (b) represents the FOV
of the \iris{} raster scan displayed in Figure \ref{fig2}(c)-(e),
and the two short vertical bars in (g) denote
that the duration of the raster scan that overlaps the splitting.
A movie for (a)--(c) is available,
which begins at 07-13T22:00 UT and ends at 07-14T06:00 UT.
The lines, arrows, and other annotations are removed in the movie.
{The real-time movie duration is $\sim$34 seconds.}}
\end{figure*}

A filament structure in solar Active Region 12665
split and erupted on 2017 July 14.
It produced a coronal mass ejection and
an M 2.4 class flare with a flux peak at 02:09 UT
{\citep{JingIL2021ApJ}}.
The Atmospheric Imaging Assembly
\citep[AIA,][]{LemenTA2012SoPh} on board {\sdo}
has observed the evolution of the filament structure.
The AIA imaging data cadence and spatial resolution are 12 s and 0\farcs6, respectively.
{Figure \ref{fig1} and \href{fig1.mp4}{Supplementary Movie}
present the imaging observations from \sdo/AIA.}
Figure \ref{fig1}(a) shows two contiguous filament branches
{(noted with ``F1'' and ``F2'' respectively)},
and Figure \ref{fig1}(b) displays a 304 \AA{} brightening between the two branches.
After the brightening, the two filament branches are separated (Figure \ref{fig1}(c)),
and their total width increases by $\sim$2\arcsec{} (Figure \ref{fig1}(g)).
{As shown in Figure \ref{fig1}(a)-(c),
the north leg of ``F2'' crosses over ``F1'' from the west
and is rooted in the south of ``F1''.
This indicates that ``F2'' is higher than ``F1'',
and otherwise their north legs will intersect.}
AIA 131 \AA{} images in Figure \ref{fig1}(d)-(e) illustrate two flares
corresponding to the two X-ray flux peaks in Figure \ref{fig1}(g).
The flares indicate two eruptions of the filament structure.
The lower branch (``F1'' in Figure \ref{fig1}(f)-(g)),
as part of the filament structure,
survives the eruptions.
These observations reveal that the filament structure is split into two branches.
{Although ``F2'' may not be exactly above ``F1'',
the filament structure resembles a ``double-decker'' geometry
\citep{LiuKT2012ApJ}.}
Afterwards, the structure undergoes a partial eruption
by ejecting the upper filament branch.

Four very large dense 400-step rasters are taken
by {\iris},
which scanned the leading polarity region of Active Region 12665
from 22:39 UT on 2017 July 13 to 00:56 UT on 2017 July 14.
Each raster consists of 400 slits
with a step size of 0\farcs35 and a pixel size of 0\farcs33.
The step cadence is 5 s with an exposure time of 4 s,
and the spectral resolution is 0.05 \AA{} for the scan.
We use the calibrated level 2 data
with dark current subtraction, flat field, and geometrical corrections applied
\citep{PontieuTL2014SoPh}.
The first of the four rasters
covers the filament splitting spatially and temporally,
which is analyzed in this study.
The scan period is 22:39 -- 23:13 UT on 2017 July 13,
and the field of view is 140\arcsec{} $\times$ 174\arcsec{}
centered at (500\arcsec, $-$150\arcsec).
To highlight the filament splitting,
only a subfield of the raster
(39\arcsec{} $\times$ 70\arcsec{} centered at (450\arcsec, $-$135\arcsec))
is investigated in detail.
The field of view of the subfield raster is marked in Figure \ref{fig1}(b).

\section{High-resolution Spectroscopic Results}\label{results}
\begin{figure*}[ht!]
\centering
\includegraphics[width=0.85\textwidth]{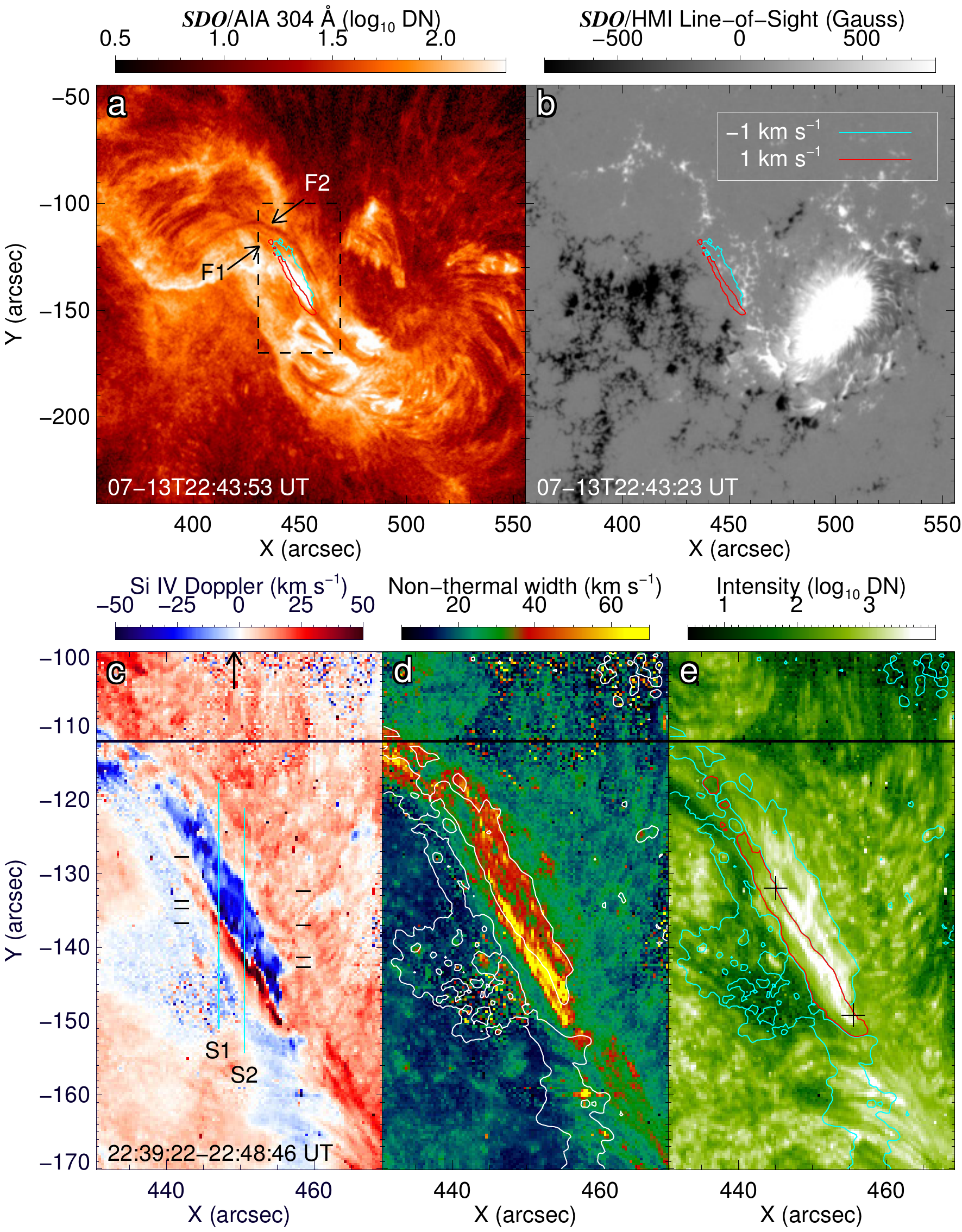}
\caption{\label{fig2}{Magnetic reconnection in the filament splitting.}
(a) \sdo/AIA 304 \AA{} image showing the brightening region
between the two filament branches.
(b) \sdo/HMI line-of-sight magnetogram of the active region.
(c)-(e) Doppler map, non-thermal width, and intensity (in digital numbers, DN) derived
from a single Gaussian fitting of \siivall{} line in the \iris{} raster scan.
Red contours in (a)-(b) and (e) are of 1 \kmps{} redshift
surrounding the brightening region.
Cyan contours in (a)-(b) and (e),
and white contours in (d) are of $-1$ \kmps{} blueshift.
The distance between the two crosses (``+'') in (e)
is $\sim$20\arcsec{} and represents the length of the reconnection region.
In (a): the rectangle denotes the field of view (FOV) of the \iris{} scan;
``F1'', ``F2'' and the arrows are similar to those in Figure \ref{fig1}(a).
In (c): the two cyan slits ``S1'' and ``S2'' indicate
where the spectra are displayed in Figure \ref{fig3}(a)-(b);
the black dashes denote where on the slits the line profiles
are plotted in Figure \ref{fig3}(c)-(d);
the time range is the duration of the \iris{} raster scan for the FOV;
the arrow points to the scan slit whose observation time
is the closest to that of the \sdo/AIA image in (a).
The horizontal line near $-110$\arcsec{} in (c)-(e) is a fiducial line.}
\end{figure*}

The filament splitting was captured on a rare occasion
with the \iris{} raster scan.
We have adjusted the coordinates of the \iris{} raster map
by cross-correlating an \iris{} \mgii{ k} wing image and an AIA 1700 \AA{} image
\citep{ChenTZ2019ScChE}.
The Doppler velocity, non-thermal width, and intensity in Figure \ref{fig2} are obtained
by fitting \siivall{} line profiles with a single Gaussian function
\citep{Peter2010AA}.
The Doppler velocity is then calibrated by removing
the Doppler shift of averaged Fe\,{\textsc{ii}} 1392.817 \AA{} line
\citep{TianZP2018ApJ}.
The non-thermal width is calculated
by subtracting the thermal and instrumental broadening
from the single Gaussian fitted width at 1/e of the peak intensity.
The above processes are described in Appendix \ref{methods}.
Neighboring large blue- and redshifts of $\gtrsim 50$ \kmps{}
in the brightening region of 304 \AA{} between the two filament branches are revealed.
For \siivall{} line,
the Doppler shifts spatially correspond to
large non-thermal widths and enhanced intensities (brightening),
and are also near the magnetic polarity inversion line
(see the \sdo{} Helioseismic and Magnetic Imager
\citep[HMI,][]{ScherrerSB2012SoPh} magnetogram in Figure \ref{fig2}(b)).
This is consistent with the observation
that a filament is located at and
parallel to a polarity inversion line
\citep[a boundary between opposite-polarity magnetic fields,][]{Martin1998SoPh}.
Doppler shift of \siiv{} line is {one of the spectroscopic signatures}
of the bidirectional outflows of magnetic reconnection
in the lower solar atmosphere
\citep{PeterTC2014Sci}.
The non-thermal broadening could be explained
by magnetic-reconnection induced turbulence
or by the velocity dispersion in the outflows
\citep[e.g.,][]{AntonucciRT1986ApJ,GordovskyyKB2016AA}.
The brightening of 304 \AA{} and \siivall{},
between the two filament branches,
indicates the hot plasma heated by magnetic reconnection
\citep{LiLN2018MNRAS,AntolinPT2021NatAs}.
The \sdo/AIA and \iris{} observations have revealed the magnetic reconnection
that occurs inside the filament structure
and splits the structure into two branches.
The length of the interface between the adjoining large blue- and redshifts,
where the reconnection site is located,
is estimated to be $\sim$20\arcsec{}
(between the two crosses in Figure \ref{fig2}(e)).
{The length is corresponding to $\sim$14 000 km on the Sun
if it is measured at the disk center.}
\iris{} took 150 seconds to scan the reconnection region.
The 304 \AA{} brightening lasts about 10 minutes,
but it is not associated with an enhancement of X-ray flux (Figure \ref{fig1}(g)).
This suggests that the reconnection in the filament splitting
releases much lower energy
than the reconnection during the filament eruption.
{The order of magnitude of the energy will be estimated in Section \ref{dem}.}

\begin{figure*}[ht!]
\centering
\includegraphics[width=0.9\textwidth]{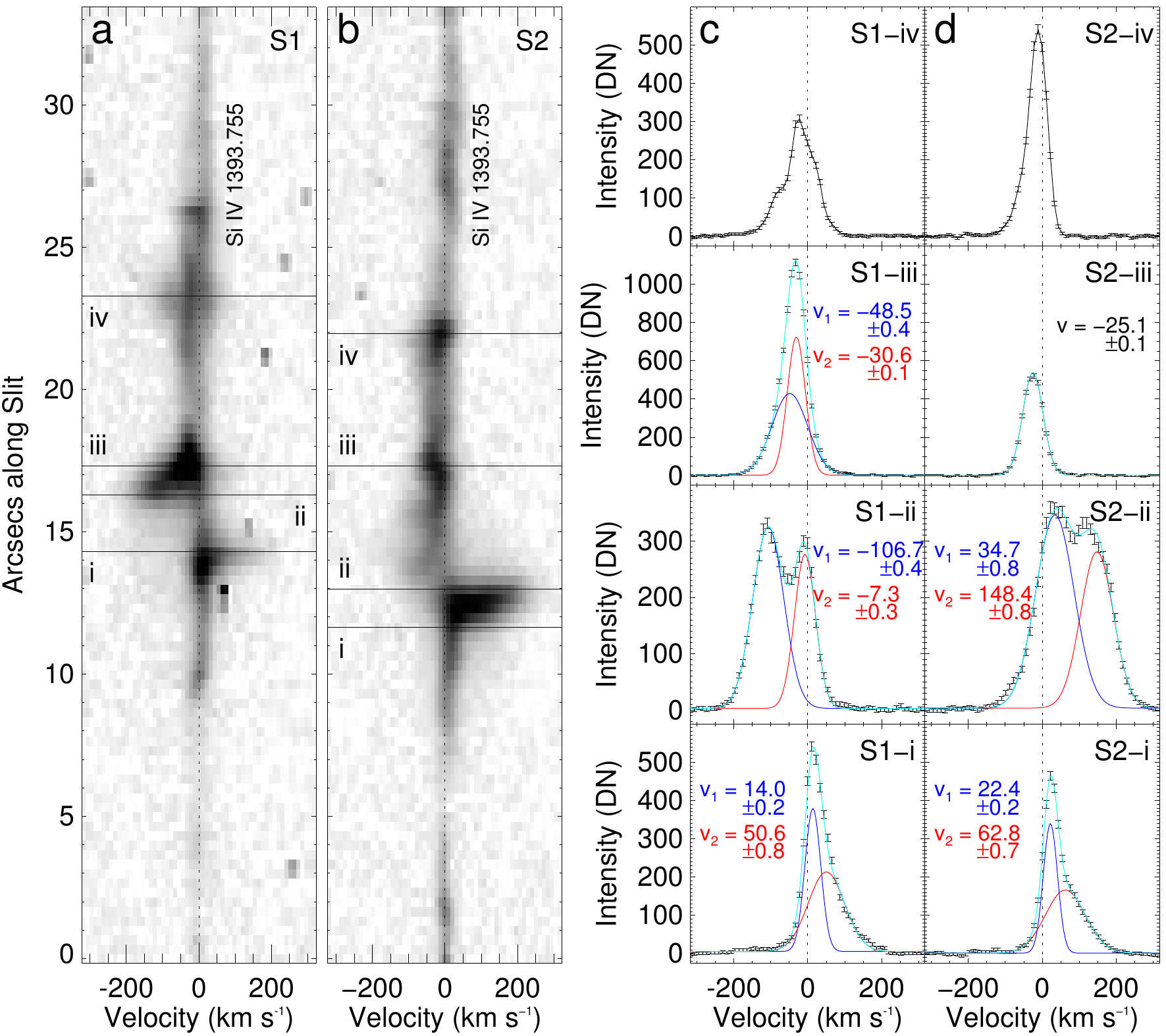}
\caption{\label{fig3}{Spectra of the bidirectional outflows
from the reconnection site.}
(a)-(b) \siivall{} line spectra on slits ``S1--2''
defined in Figure \ref{fig2}(c).
The horizontal lines ``i--iv'' on each slit
are corresponding to the dashes in Figure \ref{fig2}(c).
(c)-(d) Line profiles in positions ``i--iv'' on ``S1--2'',
and their double Gaussian fits for ``i--iii''
(but single Gaussian fit for ``S2-iii'').
The zero velocities are set to the rest wavelengths of \siiv{} \siivwav.
The black curves are the observed profiles;
the cyan curves represent the total fits;
the red and blue curves plot the two Gaussian components;
the component velocities and their 1-sigma errors are given in \kmps.}
\end{figure*}

We further investigate the spectra and profiles of \siivall{} line
on slits ``S1'' and ``S2'' specified in Figure \ref{fig2}(c).
As displayed in Figure \ref{fig3},
both blueshift and its counterpart redshift are detected on each slit,
between which the magnetic reconnection site is located.
The overall line width decreases notably on the blueshift side
away from the reconnection site on ``S1'',
and on the redshift side on ``S2'' (Figure \ref{fig3}(a)-(b)).
These indicate that the velocities projected to the line of sight
decrease after the bidirectional outflows have left the reconnection site.
Line profiles near the reconnection site are {asymmetric or double-peaked,
which} can not be fitted well with a single Gaussian function.
{A $\kappa$ distribution
can be used to fit profiles with enhanced wings
\citep[e.g.,][]{DudikPD2017ApJ}.
A double Gaussian function is also applicable
to fit the major velocity components of profiles
with asymmetric wings and especially with double peaks
\citep[e.g.,][]{Peter2010AA,HongDL2016ApJ,OrtizHN2020AA}.
In this study,}
a double Gaussian fitting (see Appendix \ref{methods})
is applied to line profiles
at ``i--iii'' on ``S1'' and at ``i--ii'' on ``S2''
(but single Gaussian fitting for ``S2-iii''; see Figure \ref{fig3}(c)-(d)).
Separate double Gaussian components are fitted
near the reconnection site on the two slits.
A blueshift component greater than 100 \kmps{} at ``S1-ii'' and
a redshift component of nearly 150 \kmps{} at ``S2-ii'' are obtained.
Beyond the reconnection site,
the blueshift decreases obviously to $\sim$48 \kmps{} at ``S1-iii'' and
the redshift is reduced noticeably to $\sim$62 \kmps{} at ``S2-i''.
These confirm that both the upward and downward outflows
decelerate remarkably after they have left the reconnection site.
We can also see line broadening,
at ``S1-iv'' and ``S2-iv'',
several arcseconds from the reconnection site on the blue wing,
which may be a signature of turbulence
\citep{JeffreyFL2018SciA,ChittaL2020ApJ}.
The turbulence can be induced
when the upward outflow interacts with the upper filament branch.
Previous simulations have demonstrated turbulence
caused by interaction of reconnection outflows
with a flux rope and/or with a flare loop top
\citep{TakahashiQS2017ApJ,ShenCR2022NatAs}.

The large Doppler shifts in the brightening region in this study
are not associated with rotational or helical motions of a filament or a jet.
The spectra of rotational or helical motions usually have a tilt pattern
because the Doppler velocity increases with the distance
from the interface between the blue- and redshifts
\citep{Rompolt1975SoPh,CurdtTK2012SoPh}.
Furthermore, the velocity of rotational or helical motions
is typically dozens of \kmps{}
\citep[e.g.,][]{DePontieuRM2014Sci,YangTP2018ApJ}.
In our case,
the spectrum manifests a zigzag but not a tilt.
Our results show that the velocity is up to the order of 100 \kmps{}
and decreases from the Doppler shift interface (Figure \ref{fig3}).

{We have also examined the other three \iris{} rasters
taken before the partial eruption.
Blueshifts of the {\siivall} line
between the two filament branches,
indicating plausible reconnection outflows,
are observed in two later rasters.
However, no conclusive counterpart redshifts
of the outflows are identified.}

\section{Analysis of differential emission measure}\label{dem}
The temperature, density, and energy
in the reconnection region are estimated
based on analysis of the differential emission measure (DEM).
The DEM is reconstructed with multiple channels of \sdo/AIA data
using the algorithm provided by
\citet{PlowmanC2020ApJ}.
The analysis procedures for the DEM
are given in Appendix \hyperref[prodem]{B}.
As displayed in Figure \ref{fig4}(a)-(b),
the increased emission measure (EM) and EM-weighted temperature ($T$)
are both along the interface between the blue- and redshifts
(i.e., the reconnection site),
which indicates that plasmas are heated there.
The temporal profiles of the DEM and $T$
averaged over four \sdo/AIA pixels near the reconnection site
are plotted in Figure \ref{fig4}(c).
$T$ reaches the peak
($\sim$$10^{7.15}$ K, 14 MK) around 22:44 UT
when the reconnection and the splitting occur (see Figure \ref{fig1}-\ref{fig2}),
and drops instantly to the pre-event level after the events.
The background temperature is $\sim$$10^{6.68}$ K (4.8 MK),
which is from averaging $\log_{10}(T)$ before the peak
(see the dashed line in Figure \ref{fig4}(c)).
The increment of $T$ during the reconnection is $\Delta T\approx 9.2$ MK.

The EM is also averaged over the four pixels
to estimate the density,
which peaks at the same time as $T$.
With a mean EM ($\sim$$4.67 \times 10^{28}$ {\pfc}) before the reconnection,
the electron density is estimated to be
$n_\mathrm{e} \approx 1.94 \times 10^{10}$ {\pcc},
by assuming a scale for the depth along the line of sight
($\sim$$1.2 \times 10^8$ cm,
see Appendix \hyperref[prodem]{B}
for the details).
With the peak EM of $\sim$$1.92 \times 10^{29}$ {\pfc},
the peak electron density
is estimated to be $\sim$$3.94 \times 10^{10}$ {\pcc}.
The thermal energy density is $E_\mathrm{th} \approx 74.7$ ergs {\pcc},
which is obtained with
$n_\mathrm{e} \approx 1.94 \times 10^{10}$ {\pcc} and $\Delta T\approx 9.2$ MK.
An ion velocity $v_\mathrm{i} \approx 100$ {\kmps} is taken
from the spectroscopic results
to calculate the kinetic energy density,
which is $E_{k} \approx 2.06$ ergs {\pcc}.
The ratio of $E_\mathrm{th}/E_\mathrm{k} \approx 36$
indicates that the kinetic energy output is ignorable in this event.
Note that the line-of-sight ion velocity given by the spectroscopy
could be $\sim$150 {\kmps} (see Figure \ref{fig3}(d)).
Because the reconnection region is $\sim$450{\arcsec}
from the disk center,
the actual velocity could be up to $\sim$200 {\kmps}
considering the projection effect and the uncertainty of the velocity direction.
If $v_\mathrm{i}$ is $\sim$200 \kmps,
$E_\mathrm{k}$ and $E_\mathrm{th}/E_\mathrm{k}$
can be $\sim$8.25 erg \pcc{} and $\sim$9, respectively.
To estimate the total increased thermal energy $W_\mathrm{th}$,
we assume that the heated plasmas are in a cylinder
lying in the reconnection region.
Then $W_\mathrm{th} \approx  1.3 \times 10^{27}$ ergs is obtained.
The total thermal energy $W_\mathrm{th}$,
the electron density (in the order of $10^{10}$ {\pcc}),
and the length of the reconnection region ($\sim$14 000 km),
are comparable to those of
transient brightenings and nanoflares
\citep[see Chapter 9 of][]{Aschwanden2005book}.
The peak electron density $\sim$$3.94 \times 10^{10}$ {\pcc}
is not used to calculate the energy,
but it does not essentially affect the order-of-magnitude estimate.
The orders of magnitude of the derived density and energy are reasonable.
However, the derived quantities are dependent
on the assumptions of spatial scales
(detailed in Appendix \hyperref[prodem]{B})
and could have significant uncertainties.
Despite the transient high temperature,
we have not seen clear Fe\,{\sc{xxi}} 1354.067 {\AA} profiles
during the reconnection.

\begin{figure}[ht!]
\includegraphics[width=0.45\textwidth]{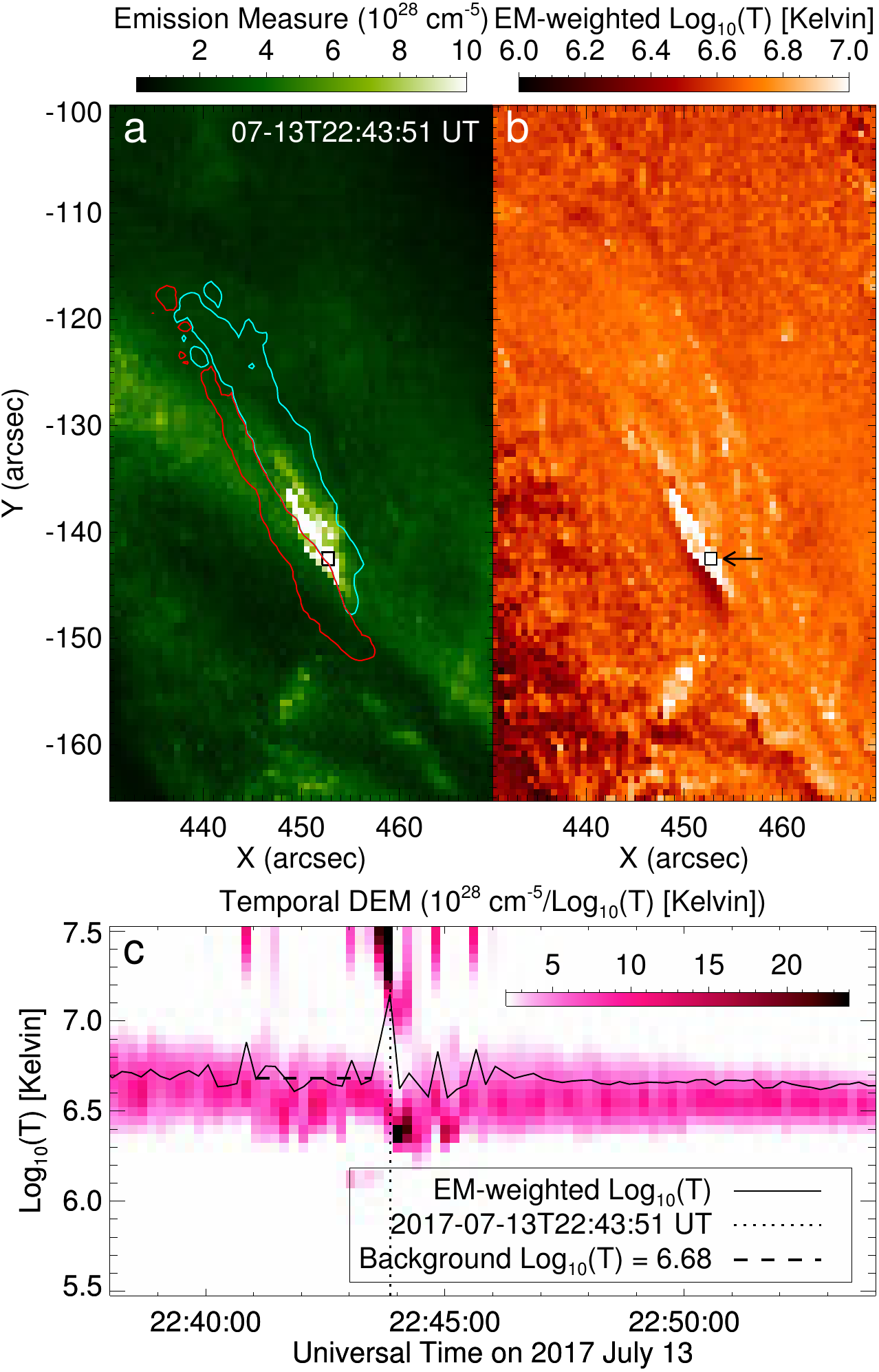}
\caption{\label{fig4}{Differential emission measure (DEM)
and emission measure (EM)
of the reconnection region based on \sdo/AIA observations.}
(a) The EM that is the zeroth moment of DEM.
(b) The EM-weighted temperature
which is the first moment of DEM divided by EM.
(c) The temporal changes in the DEM that is averaged
over the four pixels specified by the square in (b).
The field of view of (a)-(b) is similar to that of Figure \ref{fig2}(c)-(e).
The contours in (a) are the same as those in Figure \ref{fig2}(b).
In (c), the solid curve is the EM-weighted temperature corresponding to the DEM,
the dashed line denotes the averaged background EM-weighted temperature
(with its span representing the time range 22:41:03 -- 22:43:27 UT),
and the dotted line marks the peak time of the EM and EM-weighted temperature. }
\end{figure}

\section{Conclusions and Discussions}\label{conclusions}
We have investigated a magnetic reconnection
inside a solar filament structure,
using spectroscopic and imaging data from \iris{} and \sdo.
The bidirectional outflows from the reconnection site
distribute in an extended region
with a length of no less than 14 000 km on the Sun.
The velocity of the outflows are $\sim$100 \kmps{}
and decreases remarkably beyond the reconnection site.
{The temperature in the reconnection region is over 10 MK,
which is estimated based on the differential-emission-measure (DEM) analysis.}
The reconnection splits the filament structure into two upper and lower branches.
The filament structure eventually erupted partially,
with the upper branch ejected and the lower branch retained.

\begin{figure*}[ht!]
\centering
\includegraphics[width=0.8\textwidth]{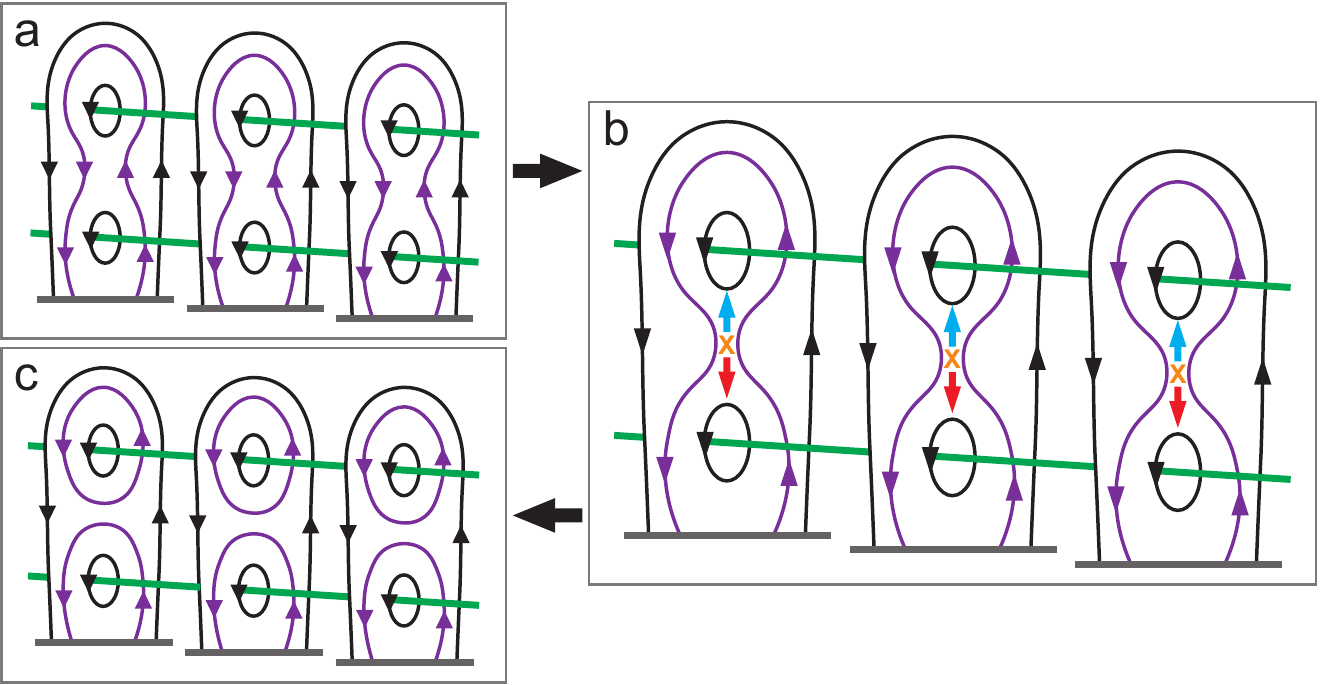}
\caption{\label{fig5}{Sketch of the magnetic field lines
associated with the reconnection in the filament splitting.}
(a) Before the reconnection the upper and lower filament branches (green lines)
are tethered by the magnetic field lines (purple curves).
(b) (enlarged view) Magnetic reconnection occurs at a series of ``X'' points
between the two filament branches,
where the blue and red arrows indicate the upward and downward outflows, respectively.
(c) After the reconnection the tether magnetic field lines are cut
and the filament structure is split.}
\end{figure*}

The unprecedented extent of the reconnection region
implies an extended reconnection of
a series of magnetic field lines in the filament structure
(see the cartoon in Figure \ref{fig5}).
The extent is comparable to
the horizontal length of the current sheet
obtained in three-dimensional magnetohydrodynamics simulations
\citep[e.g.,][]{JiangFL2021NatAs,ShenCR2022NatAs}.
Our study also shows unambiguous spectroscopic evidence
for the splitting of a filament structure by magnetic reconnection.
The reconnection in the extended region
cuts the magnetic field lines
binding the two filament branches (Figure \ref{fig5}),
after which the upper branch rises (Figure \ref{fig1}(g)).
This is reminiscent of the ``tether-cutting'' model
\citep{MooreSH2001ApJ},
and provides a mechanism for forming a ``double-decker'' filament structure
\citep{LiuKT2012ApJ}.
Interaction between the upward outflow and the upper filament branch
is inferred from the line broadening far from the reconnection site (Figure \ref{fig3}),
which may be responsible for the rise of the upper branch.
The upward reconnection outflow
is also suggested to push the upper flux rope
in a recent simulation of solar eruptions
\citep{JiangFL2021NatAs}.

{The temperature of the reconnection region
shows a transient increase to above 10 MK,
which is in the order of those of large flares,
although the total thermal energy output ($\sim$$10^{27}$ ergs)
is several orders of magnitude lower than those of large flares.
The total thermal energy dominates the kinetic energy in this event,
which is generally consistent with simulations
\citep[e.g.,][]{AunaiBS2011PhPl,ShuLL2021JGRA}.
However, energy partition in magnetic reconnection
with a similar energy level
is still in debate
\citep[e.g.,][]{InglisC2014ApJ,WarmuthM2020AA}.
}

The filament structure finally erupted partially
by ejecting the upper branch, and the lower branch remained
(see Figure \ref{fig1} and \href{fig1.mp4}{Supplementary Movie}).
The splitting and partial eruption of a filament
may contribute to successive coronal mass ejections from the same active region
\citep[e.g.,][]{BirnFH2006ApJ,ChengKD2018ApJ},
which are more likely to bring severe geomagnetic effects
than a single ejection
\citep{LiuLK2014NatCo,LugazTW2017SoPh}.
Indeed, nine days later the same active region as in our case
produced multiple filament ejections,
which caused a surprising magnetic field at 1 au as high as 68 nT
\citep{LiuZH2019ApJS}.

{This Letter has provided definite spectroscopic evidence
for the splitting of a filament structure by magnetic reconnection,
and has presented a new view on the spatial distribution
of the outflows and the thermal properties of reconnection.}

\vspace*{-0.6ex}
\section*{Acknowledgments}
We thank Dr. Yajie Chen, Dr. Leping Li, Dr. Rui Wang, and Prof. Xin Cheng,
for their valuable suggestions and discussions.
We are grateful to the anonymous referee
for constructive comments.
The research is supported by
the National Natural Science Foundation of China
(grants No. 42004145, 42274201, 42150105, and 11733003),
the Major Project of Chinese National Programs
for Fundamental Research and Development (grant No. 2021YFA0718600),
the CAS Strategic Priority Program on Space Science (grant No. XDA15018500),
and the Specialized Research Fund for State Key Laboratories of China.
H.H. is also supported by CSC (grant No. 201804910106) and MPS.
L.P.C. gratefully acknowledges funding by the European Union
(grant agreement No. 101039844).
Views and opinions expressed are however those of the author(s) only
and do not necessarily reflect those of the European Union
or the European Research Council.
Neither the European Union nor the granting authority can be held responsible for them.
\iris{} is a NASA small explorer mission developed and operated
by LMSAL with mission operations executed at NASA Ames Research Center
and major contributions to downlink communications
funded by ESA and the Norwegian Space Centre.
\sdo{} is the first mission launched for NASA's Living With a Star Program.
We acknowledge the use of data from \goes.

\facilities{\iris, \sdo, \goes}
\software{{MPFIT (\citealt{Markwardt2009MPFIT}),
          simple\_reg\_dem (\citealt{PlowmanC2020ApJ}), \&
          SolarSoftWare (\citealt{FreelandH2012ssw})}}

\appendix
\section{{Spectroscopic Analysis of {\iris} data}} \label{methods}
A single Gaussian function with a continuum,
$ I(v) = A \cdot \exp{[-\frac{(v - v_{\text{D}})^{2}}{w_{\text{1/e}}^{2}}] } + c $,
is used to fit the line profiles,
where $v$ is the given wavelength in Doppler shift units of \kmps{},
and $A$, $v_{\text{D}}$, $w_{\text{1/e}}$, and $c$ are free parameters
for the peak intensity, Doppler velocity, 1/e width, and continuum, respectively
\citep{Peter2010AA}.
The non-thermal width $w_{\text{nt}}$ in Figure \ref{fig2}(d) is determined by
$w_{\text{nt}} = \sqrt{w_{\text{1/e}}^2 -w_{\text{th}}^2 - w_{\text{instr}}^2 } $,
where $w_{\text{th}}$ is the thermal width 6.63 \kmps{}
corresponding to the formation temperature 80 000 K for \siivall{} line,
and $w_{\text{instr}}$ is the instrumental width 6.46 \kmps{}
for spectral resolution 0.05 \AA.
The total intensity in Figure \ref{fig2}(e) is calculated with
$I_{\text{tot}} = \sqrt{\pi} \cdot A \cdot w_{\text{1/e}}$.
Near the reconnection site,
some \siiv{} line profiles have two Gaussian components,
and a double Gaussian function with a continuum
is employed to fit these profiles (see Figure \ref{fig3}).
The fittings are performed using the routine \verb|mpcurvefit.pro|
in the SolarSoftWare (SSW, available at \url{https://www.lmsal.com/solarsoft})
provided by \citet{Markwardt2009MPFIT}.
The uncertainties in Figure \ref{fig3}(c)-(d)
are the 1-sigma errors given by the routine.

The Doppler velocities in this study are calibrated
by removing the Doppler shift of chromospheric Fe\,{\textsc{ii}} 1392.817 \AA{} line
\citep{TianZP2018ApJ}.
To get the Doppler shift,
the Fe\,{\textsc{ii}} line profile is averaged
over all spatial pixels in the raster,
and then is fitted by a single Gaussian function as described above.
The Doppler shift of Fe\,{\textsc{ii}} 1392.817 \AA{} line is $\sim$1 \kmps{},
which is consistent with the fact that the cold line velocity is usually trivial
\citep{PeterTC2014Sci,TianZP2018ApJ}.

An \iris{} \mgii{ k} wing image
has similar bright features to an \sdo/AIA 1700 \AA{} image
\citep{ChenTZ2019ScChE}.
We cross-correlate these two types of images
to align coordinates of \iris{} rasters to those of \sdo{} images.
An AIA 1700 \AA{} image, whose observation time
(22:43:40 UT on 2017 July 13) is during the reconnection,
is firstly cropped to the field of view of the \iris{} raster.
The \mgii{ k} wing image is made by summing the intensities
at wavelengths of 2796.352 $+$ 1.33 \AA{} and 2796.352 $-$ 1.33 \AA{},
and then resampled to the resolution of the AIA 1700 \AA{} image.
The shift between the \mgii{ k} wing and AIA 1700 \AA{} images
are computed by the routine \verb|malign.pro| in SSW.
Finally, a shift of (2\arcsec{}, 3\arcsec{}) in X and Y directions is given
and is adjusted to the coordinates of the \iris{} maps in Figure \ref{fig2}.

\section{Differential emission measure based on \sdo/AIA data} \label{prodem}
{
The differential emission measure (DEM) is obtained
using the code} {\verb|simple_reg_dem.pro|}
{in SSW provided by
\citet{PlowmanC2020ApJ}.
Six channels (94, 131, 171, 193, 211, and 335 {\AA})
of co-aligned \sdo/AIA images are the input.
The output DEM is of per unit $\log_{10} (T)$,
where $\log_{10}(T)$ is the logarithm of temperature $T$ in kelvins.
The emission measure (EM) is given by
$\mathrm{EM} = \int \mathrm{DEM}(T_\mathrm{log}) dT_\mathrm{log}$,
where $T_\mathrm{log} = \log_{10} (T)$ ranges from 5.5 to 7.5
with a step of 0.05.
The logarithm of EM-weighted temperature is
$\overline{T}_\mathrm{log}
= \int \mathrm{DEM}(T_\mathrm{log}) T_\mathrm{log} dT_\mathrm{log}/\mathrm{EM}$
\citep[for details see][]{PlowmanC2020ApJ}.}

{
The electron density is $n_\mathrm{e} = \sqrt{\mathrm{EM}/l}$,
where $l$ is the depth along the line of sight.
The depth $l \approx 1.2 \times 10^8$ cm
is assumed to be equivalent to
the width of the region with enhanced EM and EM-weighted temperature
(see Figure \ref{fig4}(a)-(b)).
The thermal energy density is obtained with
$E_\mathrm{th} = 3 n_\mathrm{e} k_\mathrm{B} \Delta T$,
where $k_\mathrm{B}$ is the Boltzmann constant
and $\Delta T$ is the increased EM-weighted temperature.
The kinetic energy density is calculated with
$E_\mathrm{k} = \frac{1}{2} \mu m_\mathrm{H} n_\mathrm{e} v_\mathrm{i}^2$,
where $\mu \approx 1.27$ is the mean molecular weight
\citep[see Chapter 3 of][]{Aschwanden2005book},
$m_\mathrm{H}$ is the hydrogen mass,
and $v_\mathrm{i} \approx 100$ \kmps{}
is the ion velocity estimated from the \siiv{} line Doppler shift.
Assuming that the heated plasmas
are in a cylinder volume $V$,
the total thermal energy is given by
$W_\mathrm{th} = E_\mathrm{th} V$.
The cylinder height and diameter
are $\sim$$1.4 \times 10^9$ cm
(the distance between the two crosses (``+'') in Figure \ref{fig2}(e))
and $\sim$$1.2 \times 10^8$ cm
(the depth along the line of sight), respectively.
}

\end{CJK*}
\end{document}